\documentclass[prd,twocolumn,superscriptaddress,altaffilletter,amssymb,showpacs,nofootinbib]{revtex4}
\usepackage[dvips]{graphicx}
\usepackage{amsmath}
\newcommand{\be}{\begin{equation}}
\newcommand{\ee}{\end{equation}}
\newcommand{\bea}{\begin{eqnarray}}
\newcommand{\eea}{\end{eqnarray}}

\begin{document}

\title{Time-like vs Space-like Extra Dimensions}

\author{Israel Quiros}\email{israel@uclv.edu.cu}
\affiliation{Universidad Central de Las Villas, Santa Clara CP 54830, Cuba}

\date{\today}
\begin{abstract}
Higher-dimensional theories with time-like and space-like extra dimensions are compared both from the conceptual and from the phenomenological points of view. In this context causality and unitarity are discussed. It is shown that additional time-like dimensions allow to recover four-dimensional phenomenology without invoking neither Kaluza-Klein compactification procedure nor brane world construct. It is shown, also, that contrary to scenarios with space-like extra dimensions, in higher-dimensional space-times with additional time-like dimensions the cosmological constant problem can be safely solved. 
\end{abstract}


\maketitle

\section{Introduction}

The standard tenet in dealing with higher-dimensional theories is to consider space-like extra dimensions almost exclusively. Time-like extra dimensions have been disregarded due to serious conflicts with causality and unitarity \cite{yndurain,dvali,erdem}. It can be shown, however, that the reasoning line behind violations of causality and of unitarity in space-times with additional time-like dimensions, has been prejudiced by considerations that are deeply rooted in the belief that space-like and time-like extra dimensions have to be considered on an equal footing.

The aim of the present study is to show that the latter belief is in general not justified, and that there exists an alternative approach to higher-dimensional theories with time-like extra dimensions that is not in conflict neither with causality nor with unitarity. The fact that, under this alternative interpretation, four-dimensional phenomenology may be recovered without invoking neither Kaluza-Klein (KK) compactification nor brane confinement mechanisms, will be particularly discussed. It will be shown, also, that the cosmological constant problem may be solved in higher-dimensional theories with time-like extra dimensions. In this regard, additional space-like dimensions are not of any help.

\section{Causality and Unitarity}

In the present study, for sake of simplicity, gravity effects will not be considered. We shall consider five-dimensional (5D) space-times with the following flat metric: $\eta_{AB}=(\eta_{\mu\nu},\epsilon)$, where $\eta_{\mu\nu}=(-1,\delta_{ik})$ is the four-dimensional (4D) flat Minkowski metric. The symbol $\epsilon=\pm 1$ ($+1$ for a space-like extra dimension, $-1$ for a time-like one). Capital Latin indexes $A,B,...=0,1,2,3,5$, while Greek indexes run through ordinary 4D space-time ($\alpha,\beta,...=0,1,2,3$). Small Latin indexes run through 3D space ($i,j,...=1,2,3$). We shall assume a compact extra dimension that is spanned by the fifth coordinate $y\equiv x^5$. In consequence the field variables will be periodic in $y$: $\Phi(x,y)=\Phi(x,y+2\pi L)$, where $x\equiv\{x^\alpha\}$ and $L$ is the size of the compact extra dimension. Large extra dimensions may be recovered in the continuous limit $L\rightarrow\infty$.

It will be instructive to study the theory of a real scalar field $\psi$ of mass $m$, that obeys the 5D Klein-Gordon (KG) equation:

\be (\Box_{(5)}-m^2)\psi=0,\label{kg}\ee where $\Box_{(5)}\equiv\eta^{NM}\partial_N\partial_M=\eta^{\mu\nu}\partial_\mu\partial_\nu+\epsilon\partial_y^2$. This equation can be written in a manifestly 4D form if one considers the following decomposition:

\be \psi(x,y)=\sum_n \psi_n(x)\;e^{iny/L},\;\;n=0,\pm1,\pm2,...,\label{decomposition}\ee that is dictated by the assumed topology of the extra dimension. After the decomposition (\ref{decomposition}), the KG equation (\ref{kg}) can be written in a form that is adequated to a 4D observer like us:

\be \{\bar\Box-(m^2+\epsilon\; n^2/L^2)\}\psi_n(x)=0,\label{4dkg}\ee where $\bar\Box\equiv\eta^{\mu\nu}\partial_\mu\partial_\nu$ is the standard 4D flat D'Lambertian. As a consequence of the additional dimension, in addition to the fundamental mode ($n=0$), there is a tower of KK excitations that are seen by a 4D observer as a discretum of different particles with masses $m_n^2=m^2+\epsilon\; n^2/L^2$. For a space-like extra dimension ($\epsilon=+1$) the mass squared is always positive definite, however, if the additional dimension is time-like ($\epsilon=-1$), it is not in general positive definite. In the latter case scalar field modes with $|n|> mL$ are seen by a 4D observer as tachyonic degrees of freedom that originate violations of 4D causality and of unitarity.

There exists however an alternative approach that is free of tachyons and, consequently, does not conflict neither with causality nor with unitarity. To show this, let us consider the plane-wave approach: $\psi_n(x)=\psi_n^0\exp{[-i(E_n t-{\bf p}_n{\bf x})]}$, where $E_n,\;{\bf p}_n$ are the energy and spatial momentum of the $n$-th scalar field excitation respectively. Equation (\ref{4dkg}) then leads to the following energy-momentum relationship:

\be E_n^2-{\bf p}_n^2-m^2-\epsilon\frac{n^2}{L^2}=0.\label{energymomentum}\ee

Consider the case for a time-like extra dimension ($\epsilon=-1$). It is evident from (\ref{energymomentum}) that a second alternative interpretation is indeed possible. Instead of grouping the term coming from twice $y$-derivative (term $n^2/L^2$ in (\ref{energymomentum})) under mass squared: $m_n^2=m^2-n^2/L^2$, it is perhaps more appropriate to group it with the energy squared term (coming from twice $t$-derivative) to get a modified energy squared:

\be \bar E_n^2=E_n^2+n^2/L^2.\label{modifiedenergy}\ee

According to this alternative picture, the scalar field excitations of the same mass that come from the time-like extra dimension, differ from the pure 4D mode ($n=0$) just in their energies. I. e., unlike standard KK modes that represent different particles, these can be viewed as excited energy states of a same particle.

It is evident that, according to this tachyonless alternative approach, there is no any conflict between an additional (compact) time-like dimension and 4D causality. There is no any conflict with unitarity neither. Actually, according to the standard interpretation, since the following energy-momentum relationship takes place:

\be E_n^2-{\bf p}_n^2=m_n^2\equiv m^2+\epsilon\frac{n^2}{L^2},\label{standardem}\ee in the rest frame one has $\psi_n(t)\propto\exp{(im_n t)}$, so that, given a time-like extra dimension ($\epsilon=-1$), and provided that $|n|>Lm$, then: $\psi_n(t)\propto\exp{(-\sqrt{n^2/L^2-m^2}\;t)}$. This means that, since the amplitude squared of the given tachyonic mode $|\psi_n|^2\propto\exp{(-2\sqrt{n^2/L^2-m^2}\;t)}$, the corresponding KK excitations are unstable and decay into "nothingness" in a time $\tau_n\sim L/\sqrt{n^2-L^2m^2}$.

Unlike this, according to the tachyonless perspective, since the energy-momentum relationship (\ref{standardem}) gets replaced by

\be \bar E_n^2-{\bf p}_n^2=m^2,\label{modifiedem}\ee where, for a time-like extra dimension the modified energy squared $\bar E_n^2$ is given by (\ref{modifiedenergy}), then, in the rest frame $\psi_n(t)\propto\exp{(im t)}\rightarrow |\psi_n|^2\propto 1$. The corresponding modes are stable and unitarity is preserved.

It is apparent that the above arguments may be safely used to extend the discussion to more than one additional time-like dimension.

\section{Phenomenology}

When extra dimensions are considered, one has to wonder whether the given higher-dimensional theory appropriately describes 4D phenomenology. In particular one has to care about recovering of standard Newton's law of gravity. For an extra dimension, the static weak-field limit of general relativity yields to the following modified Poisson's equation for the gravitational potential:\footnotemark\footnotetext{This is usually done by looking for small perturbations around the flat background. In the present case one has to replace the 5D flat metric $\eta_{AB}$ by the perturbed one $\eta_{AB}+h_{AB}$, where $h_{AB}(x,y)$ are the small perturbations. The transverse, traceless gauge ($\partial^Nh_{NA}=h^N_N=0$) is usually considered. In the static case, the Newtonian gravitational potential is just: $V=h_{00}/2$.}

\be ({\bf\nabla}^2+\epsilon\partial_y^2)V({\bf x},y)=4\pi\rho({\bf x},y),\label{poisson}\ee where $\rho$ is the mass distribution generating the static weak gravitational field around it, and ${\bf\nabla}^2\equiv\delta^{ik}\partial_i\partial_k$ is the customary 3D (flat) Laplace's operator. When the extra dimension is space-like, solutions to (\ref{poisson}) in empty space yield to known modified forms of the Newtonian potential. In standard KK theories with additional spatial dimensions 4D phenomenology is recovered by invoking the so called KK compactification procedure: massive KK modes that produce modifications to standard 4D graviton (fundamental mode $n=0$) propagation, lie behind energies so far attained in experimental tests of Newton's law.

If, as it is usually done, one were to treat time-like and space-like extra dimensions on an equal footing then, consideration of a time-like extra dimension in equation (\ref{poisson}) would yield to imaginary contributions to the gravitational self-energy of objects. This fact imposes stringent phenomenological bounds to considering additional time-like dimensions as a serious alternative \cite{dvali,matsuda}.\footnotemark\footnotetext{In addition there are serious phenomenological bounds imposed by nucleon decay phenomenology \cite{yndurain,dvali,erdem}. Notwithstanding, as it will be shown later on, the corresponding analysis, based on the study of particle propagators, can be made compatible with the alternative picture explained in the present study.}

However, as already discussed in the precedent section, there is an alternative approach to time-like extra dimensions that is consistent with 4D causality and unitarity. It will be inmediately shown that this alternative picture is consistent also with 4D phenomenology. For the latter purpose it will be useful to realize that the notion of "staticity" entails an obvious separation of the notions of space and time. In fact, a given configuration $C$ is said to be static if the corresponding field variables $\{C_i\}$ are independent of time. If space-like extra dimensions are being considered, the aforementioned notion of "static configuration" does not need to be modified. However, as long as additional time-like dimensions are concerned, this notion has to be necessarily modified. Actually, as before, let us consider just a time-like extra dimension spanned by the additional time-coordinate $y\equiv x^5$ (the usual time-coordinate is spanned by $t\equiv x^0$). It is evident now that, if the tachyonless approach to time-like extra dimensions is undertaken, then a given configuration $C$ is said to be static whenever the field variables $\{C_i\}$ are independent of both time-coordinates $t$ and $y$. Therefore, the addition of (any number of) time-like extra dimensions does not modify the standard 3D Poisson's equation for the Newtonian gravitational potential, generated by static distributions of mass (once again recall that static now means independence of both $t$ and $y$). Equation (\ref{poisson}) has to be, accordingly, replaced by:

\be {\bf\nabla}^2V({\bf x})=4\pi\rho({\bf x}).\label{3dpoisson}\ee I. e., Newton's law of gravity is not modified in any way by the presence of time-like extra dimensions.

It is evident that, according to the present approach to time-like extra dimensions, any field degrees of freedom living in the 5D space-time are constrained to propagate in the common to all particles 3D Euclidean space with metric tensor $\delta_{ik}$. Therefore, neither KK compactification nor any other confining mechanizm are necessary to recover 4D phenomenology.

Regarding other phenomenological bounds usually considered, such as, for instance, nucleon decay rate \cite{dvali,yndurain,erdem}, it has to be pointed out that the feasibility of these bounds rests on the usual approach to additional (compact) time-like dimensions. Due to the appearance of tachyon modes in that approach, the particle propagators are modified. For example, the propagator of a photon would become:

\be D_{\mu\nu}=-i\eta_{\mu\nu}\frac{1}{p^2-\epsilon(n^2/L^2)+i0},\label{photonpropagator}\ee where $p=\{p^\mu\}$ are the momenta conjugated to the ordinary 4D space-time variables: $p^2=E^2-{\bf p}^2$. For a time-like extra dimension ($\epsilon=-1$) there will be unphysical poles at $p=\pm in/L$, that will modify, for instance, the transition amplitudes for electromagnetic processes in atomic nuclei. There will appear, in particular, imaginary contributions to these amplitudes, meaning, in turn, unphysical decay processes that are not seen in experiments. On the contrary, according to our tachyonless approach, the 4D propagator (\ref{photonpropagator}) can be written, alternatively, in the following way:

\be \bar D_{\mu\nu}=-i\eta_{\mu\nu}\frac{1}{\bar p^2+i0},\label{modifiedpropagator}\ee where $\bar p^2=\bar E^2-{\bf p}^2$, with $\bar E^2$ given by $\bar E^2=E^2-\epsilon(n^2/L^2)$ (compare with equation (\ref{modifiedenergy}) for a time-like extra dimension). According to this alternative approach, the standard photon propagator is not modified by the presence of the extra dimension, instead, there appears a tower of excited energy states of the massless 4D photon. It is evident from this discussion that the amplitude of given electromagnetic processes are not modified by the presence of time-like extra dimensions. In consequence, the given phenomenological bounds are out of place and have to be abandoned.

Before concluding this section we want to underline that, while according to the standard approach to additional dimensions, the KK compactification procedure (or alternative mechanizms like brane world construct) has to be applied to adequately reproduce 4D phenomenology, if our alternative (tachyonless) approach is correct, the addition of (compact) time-like extra dimensions does not require of any specially invented mechanizm to be phenomenologically viable. In this case, instead of a tower of (infinitely many) additional particles of different masses, one has a collection of excited energy states (an energy discretum quantized in units of $1/L$) of a same particle.

\section{Extra dimensions and Vacuum Energy}

Since long ago it is known that zero-point fluctuations of the fields contribute towards a non-null (perhaps infinite) energy density of empty space. While field theoretic calculations yield to a hughe value for it, that is set roughly by the Planck energy scale $\left\langle \rho_v\right\rangle\sim 10^{71} GeV^4$, the cosmological observations point to a tiny (not yet null) value $\left\langle \rho_v\right\rangle_{obs}\sim 10^{-47} GeV^4$. This catastrophic discrepancy between theoretical predictions and observational data is known as the "cosmological constant problem" (CCP) (see, for instance, the reviews \cite{weinberg,sahni}). 

It will be very instructive to see how the presence of an additional space-time dimension can modify the standard 4D field theoretic computation of vacuum energy density. A crude estimate can be obtained just by replacing the usual formula

\be \left\langle \rho_v\right\rangle=\frac{1}{(2\pi)^3}\int_0^\Lambda d^3p \frac{1}{2}\sqrt{p^2+m^2},\ee where $\Lambda$ is some momentum cutoff of the order of the Planck mass, by the following one:

\be \left\langle \bar\rho_v\right\rangle=\sum_n\frac{1}{(2\pi)^3}\int_0^\Lambda d^3p \frac{1}{2}\sqrt{p^2+m^2+\epsilon\frac{n^2}{L^2}}.\label{modifiedvacuum}\ee 

The case for a space-like extra dimension ($\epsilon=+1$) has to be discussed separately from the case for an additional time-like dimension ($\epsilon=-1$). In the first case, since the standard KK compactification procedure has to be applied in order to recover 4D phenomenology, then the leading contribution to vacuum energy density is given by the fundamental modes with $n=0$, so that (see in \cite{weinberg}):

\be \left\langle\bar\rho_v\right\rangle\approx\left\langle\rho_v\right\rangle\approx\frac{\Lambda^4}{16\pi^2}=2\times 10^{71} GeV^4,\label{computation}\ee where, as customary, it has been assumed that $\Lambda\gg m$. Otherwise, 4D phenomenology prevents additional space-like dimensions to modify in any fundamental way the CCP.

On the contrary, according to our tachyonless approach to additional space-time dimensions, the presence of time-like extra dimensions does not entail conflicts with 4D phenomenology, even if the size of the extra dimension is large. Actually, as already discussed, if we adhere to the alternative interpretation of time-like extra dimensions undertaken here, the most stringent phenomenological bounds that are set by violations of causality and of unitarity \cite{dvali,yndurain,erdem} are not valid any more. Therefore one may turn to the continuous limit $L\rightarrow\infty$ (large extra dimension) without modifying the above picture. In this case the symbol $\sum$ in equation (\ref{modifiedvacuum}) may be replaced by $L\int d\xi$, where $n/L\rightarrow\xi$, i.e.,

\be \left\langle\bar\rho_v\right\rangle\rightarrow\frac{L}{(2\pi)^3}\int_0^\Lambda d^3p \int_{-\infty}^\infty d\xi \frac{1}{2}\sqrt{p^2+m^2-\xi^2}.\ee Besides, since under the integral there is an even function of $\xi$, then $\int_{-\infty}^\infty\rightarrow 2\int_0^\infty$. Therefore, in the continuous limit, instead of (\ref{modifiedvacuum}), the energy density of empty space may be computed with the help of the following equation:

\be \left\langle\hat\rho_v\right\rangle=\frac{2L}{(2\pi)^3}\int_0^\Lambda d^3p \int_0^\lambda
d\xi \frac{1}{2}\sqrt{p^2+m^2-\xi^2},\label{continousvacuum}\ee where a certain "momentum" cutoff $\lambda$ has been considered. Mathematically the latter cutoff is imposed by requiring the vacuum energy density to be a real quantity. Integration in $\xi$-variable yields to the following expression:

\bea &&\left\langle\hat\rho_v\right\rangle=\frac{L}{(2\pi)^3}\int_0^\Lambda d^3p\frac{1}{2}\{\lambda\sqrt{p^2+m^2-\lambda^2}+\nonumber\\
&&(p^2+m^2)\arcsin{(\frac{\lambda}{\sqrt{p^2+m^2}})}\}.\label{equation}\eea

In order for the latter integral to be defined one has to require that $\sqrt{p^2+m^2}\geq\lambda$. This means that, while $\Lambda$ represents some ultraviolet cutoff in momentum space, $\lambda$ represents an infrared cutoff instead. In the limit when $\Lambda\gg\lambda$, the integral in the right-hand-side of equation (\ref{equation}) can be easily computed (compare with (\ref{computation})):

\be \left\langle\hat\rho_v\right\rangle\approx\frac{L\lambda}{(2\pi)^3}\int_0^\Lambda d^3p\sqrt{p^2+m^2}\approx\frac{L\lambda\Lambda^4}{8\pi^2}.\ee

It is surprising that, after applying the aforementioned procedure of going to the continuous limit $L\rightarrow\infty$, the catastrophic disagreement between vacuum energy density computation and observations can be safely removed just by requiring that $L\lambda\approx 0.25\times 10^{-118}$. It is important to recall that the continuous limit can not be applied to the case with space-like extra dimensions due to conflicts with 4D phenomenology arising from appearance of a continuum of infinitely light KK modes.

Worth mentioning that, as already noted, together with the ultraviolet cutoff set by $\Lambda$, there appears also an infrared cutoff given by $\lambda\sim 10^{-119}L^{-1}$. In this context, the fact that the energy density of empty space, although vanishingly small, is not yet exactly null, means that the infrared cutoff indeed exists and plays a role. Of course, the latter statement is strongly dependent on the validity of the tachyonless approach to time-like extra dimensions discussed here.

\section{Concluding remarks}

We have shown that there can be an alternative approach to time-like (compact) extra dimensions that is free of tachyons and, consequently, does not conflict neither with causality nor with unitarity. According to this approach, the standard picture with an infinite tower of Kaluza-Klein particles of different masses has to be replaced by one in which there is a discretum of excited energy states of a same particle instead. 

If our alternative approach is correct, addition of time-like extra dimensions does not entail conflicts with 4D phenomenology neither. In this case standard phenomenological bounds to time-like extra dimensions are not applicable any more. The consequence is that one may turn to the continuous limit ($L\rightarrow\infty$) and the contribution of the extra dimension to the energy density of empty space can be easily computed. The surprise is that, thanks to apearance of an infrared cutoff $\lambda\sim 10^{-119}L^{-1}$, the cosmological constant problem can be safely solved.

\acknowledgements The author thanks the MES of Cuba by partial financial support of the present research.

\end{document}